\documentclass{iau}
\usepackage{graphicx}
\def\be{\begin{equation}}
\def\ee{\end{equation}}
\def\bary{\begin{eqnarray}}
\def\eary{\end{eqnarray}}
\def\en{E_\nu}

\title[Lepto-hadronic processes and energy-high neutrinos in NGC 1275] 
{Lepto-hadronic processes and high-energy neutrinos in NGC 1275}

\author[N. Fraija et al.]   
{N.~Fraija,$^1$\thanks{Luc Binette-Fundaci\'on UNAM Fellow.} A.~Marinelli,$^2$ U.~Luviano-Valenzuela,$^{1,3}$ A.~Galv\'an-Gam\'ez,$^{1,3}$ \and C.~ Peterson-B\'orquez$^{1,3}$}

\affiliation{$^1$Instituto de Astronom\' ia, Universidad Nacional Aut\'onoma de M\'exico, Circuito Exterior,
email: {\tt nifraija@astro.unam.mx} \\[\affilskip]
$^2$Instituto de Fisica, UNAM, Circuito Exterior, email: {\tt antonio.marinelli@fisica.unam.mx} \\[\affilskip]
$^3$Facultad de Ciencias, UNAM, Circuito Exterior, email: {\tt  urieluval@ciencias.unam.mx,  g.antonio@ciencias.unam.mx and christian.pb.94@ciencias.unam.mx} }
\pubyear{2014}
\volume{313}  
\pagerange{119--126}
\jname{Extragalactic jet from every angle}
\editors{F.Massaro, C.C.Cheung, E.Lopez \& A.Siemiginowska, eds.}
\begin{document}
\maketitle
\begin{abstract}
The nearby active galaxy NGC 1275,  has widely been detected from radio to gamma rays.  Its spectral energy distribution (SED) shows a double-peak feature, which is well explained by synchrotron self-Compton (SSC) model. However, recent TeV detections might suggest that very-high-energy $\gamma$-rays (E$\geq$100 GeV) may not have a leptonic origin. We test a lepto-hadronic model to describe the whole SED through SSC emission and neutral pion decay resulting from p$\gamma$ interactions. Also, we estimate the neutrino events expected in a Km$^3$ Cherenkov telescope.

\end{abstract}
\firstsection 
\section{Introduction}
NGC 1275, also known as Perseus A and 3C 84, is the nearby active galaxy located at the center of the Perseus cluster at  redshift of $z=0.0179$  (\cite[Veron (1978)]{1978Natur.272..430V}).  This object has been well studied in radio,  optical/UV,  X-ray and  MeV - GeV emission (\cite[Abdo \etal\ (2009)]{2009ApJ...699...31A}, \cite[Aleksi\'c \etal\ (2012)]{2012A&A...539L...2A}, \cite[Acciari \etal\ (2009)]{2009ApJ...706L.275A}).   Recently, this source has been detected by MAGIC telescopes with a statistical significance of $6.6\,\sigma$ above 100 GeV in 46 hr of stereo observations carried out between August 2010 and February 2011.  The measured differential energy spectrum between 70 GeV and 500 GeV can be described by a power law with a steep spectral index of {\small $\Gamma=-4.1\pm0.7_{stat}\pm0.3_{syst}$}, and the average flux above 100 GeV is {\small $F_{\gamma}=1.3\pm 0.2_{stat}\pm0.3_{syst}\times 10^{-11}\,cm^{-2}\,s^{-1}$} (\cite[Aleksi\'c \etal\ (2012)]{2012A&A...539L...2A]}).  We propose a lepto-hadronic model to describe the whole SED as the superposition of synchrotron self-Compton (SSC) emission and the neutral pion decay products.

%
%
\section{Theoretical Model}
Fermi-accelerated electrons in the emitting region are described by a broken power-law written as: {\small $N_e(\gamma_e)   \propto \gamma_e^{-\alpha_e}$ for  $\gamma_{e,m}<\gamma_e < \gamma_{e,b}$, and $N_e(\gamma_e)\propto\gamma_{e,b}    \gamma_e^{-(\alpha_e+1)} $ for $\gamma_{e,b} \leq  \gamma_e<\gamma_{e,max}$} and co-accelerated protons at the same place for a simple power law as: {\small $N_p(\gamma_p) \propto \gamma_p^{-\alpha_p}$}, where $\alpha_e (\alpha_p)$ is  the power index for electrons (protons) and {\small $\gamma_{e,i}$} ({\small $\gamma_p$}) are the electron (proton) Lorentz  factors. The index $i$, $c$ and $max$ are for minimum, break and maximum, respectively.  Assuming that the emitting region is endowed by a magnetic field, electrons and protons are cooled down by electromagnetic (synchrotron radiation and Compton scattering) and hadronic (proton-photon interaction) channels, respectively.  Taking into account the cooling processes for electron and proton distributions,  the observed spectra for electron synchrotron radiation, Compton scattering and proton-photon interaction  are (\cite[Fraija  (2014a)]{2014MNRAS.441.1209F}, \cite[Fraija \etal\ (2012)]{2012ApJ...753...40F})
{\small
\begin{eqnarray}
\label{espsyn}
 \left(\epsilon^2 N(\epsilon)\right)_{\gamma,syn} = A_{syn,\gamma} \cases {
(\frac{\epsilon_\gamma}{\epsilon^{syn}_{\gamma,m}})^{4/3}                                          &  $\epsilon_\gamma < \epsilon^{syn}_{\gamma,m}$                                                        \cr  
 (\frac{\epsilon_\gamma}{\epsilon^{syn}_{\gamma,m}})^{-(\alpha-3)/2}                          &  $\epsilon^{syn}_{\gamma,m} < \epsilon_\gamma < \epsilon^{syn}_{\gamma,c}$                                                             \cr  
(\frac{\epsilon^{syn}_{\gamma,c}}{\epsilon^{syn}_{\gamma,m}})^{-(\alpha-3)/2}    (\frac{\epsilon_\gamma}{\epsilon^{syn}_{\gamma,c}})^{-(\alpha-2)/2},   &  $\epsilon^{syn}_{\gamma,c} < \epsilon_\gamma < \epsilon^{syn}_{\gamma,max}$\,,        \cr  
}
\end{eqnarray}
}
{\small
\begin{eqnarray}
\label{espsyn}
 \left(\epsilon^2 N(\epsilon)\right)_{\gamma,ssc}= A_{ssc,\gamma} \cases {
(\frac{\epsilon_\gamma}{\epsilon^{ssc}_{\gamma,m}})^{4/3}    &  $\epsilon_\gamma < \epsilon^{ssc}_{\gamma,m}$,\cr
 (\frac{\epsilon_\gamma}{\epsilon^{ssc}_{\gamma,m}})^{-(\alpha-3)/2}  &  $\epsilon^{ssc}_{\gamma,m} < \epsilon_\gamma < \epsilon^{ssc}_{\gamma,c}$,\cr
(\frac{\epsilon^{ssc}_{\gamma,c}}{\epsilon^{ssc}_{\gamma,m}})^{-(\alpha-3)/2}    (\frac{\epsilon_\gamma}{\epsilon^{ssc}_{\gamma,c}})^{-(\alpha-2)/2}           &  $\epsilon^{ssc}_{\gamma,c} < \epsilon_\gamma < \epsilon^{ssc}_{\gamma,max} $\,,\cr
}
\end{eqnarray}
}
and 
{\small
\begin{eqnarray}
\label{pgamma}
\left(\epsilon^2 N(\epsilon)\right)_{\pi^0, \gamma} = A_{p,\gamma} \cases{
\left(\frac{\epsilon_{\gamma}}{\epsilon_{0}}\right)^{-1} \left(\frac{\epsilon_{\gamma,c,\pi^0}}{\epsilon_{0}}\right)^{-\alpha_p+3}          &   $\epsilon_{\gamma} < \epsilon_{\gamma,c,\pi^0}$\cr
\left(\frac{\epsilon_{\gamma}}{\epsilon_{0}}\right)^{-\alpha_p+2}                                                                                        &   $\epsilon_{\gamma,c,\pi^0} < \epsilon_{\gamma}$\,,
}
\end{eqnarray}
}
\noindent 
{\small respectively, where  $A_{syn,\gamma}$ ($A_{ssc,\gamma}$), $\epsilon^{syn}_{\gamma,m}$ ($\epsilon^{ssc}_{\gamma,m}$), $\epsilon^{syn}_{\gamma,c}$ ($\epsilon^{ssc}_{\gamma,c}$) and $\epsilon^{syn}_{\gamma,max}$ ($\epsilon^{ssc}_{\gamma,max}$)  are the proportionality constant and break energies for characteristic, cut-off and maximum of electron synchrotron (Compton scattering) spectrum and $A_{p,\gamma}$ and $\epsilon_{\gamma,c,\pi^0}$ are the proportionality constant and break photon energy for photo-pion spectrum} (\cite[Fraija  (2014b)]{2014ApJ...783...44F}).  It is important to highlight that  the neutrino counterpart  is calculated through {\small $\int \frac{dN_{\nu}}{d\en}\,\en\,d\en=\frac14\int \left(\frac{dN}{d\epsilon}\right)_{\pi^0,\gamma}\,\epsilon_{\pi^0,\gamma}\,d\epsilon_{\pi^0,\gamma}$}, where the neutrino flux is  {\small  dN$_\nu/d\en=A_{\nu} \,\en^{-\alpha_\nu}$}, with $\alpha_\nu\simeq\alpha_p$.
\section{Results}
{\small We have presented a lepton-hadronic model through eqs. \ref{espsyn}, \ref{espsyn} and \ref{pgamma} to describe the broadband SED of NGC1275. For the hadronic interactions we evoke the p$\gamma$ interactions of  Fermi-accelerated protons with the target photons at the second SSC peak.  To find the best fit of our leptonic and hadronic models, we use the method of chi-square ($\chi^2$) minimization.  The fit values for the leptonic model are shown in Table 1 and 2 while  for hadronic model are  {\small $A_{p\gamma}=(2.932\pm 0.6579) \times10^{-6}\, {\rm MeV\,cm^{-2}\, s^{-1}}$} and {\small  $\alpha_p=3.467 \pm 0.090$}.  The leptonic model describes the photon spectrum up to a few GeVs while the hadronic model up to hundreds of GeVs. The neutrino event expected in a Km$^3$ Cherenkov telescope  is $0.93\times 10^{-4}$ per year.} \\ 
\begin{tabular}{ll}
\begin{tabular}{cc}
 \hline
 \scriptsize{Parameter} & \scriptsize{Value}\\
 \hline 

 \scriptsize{$A_{syn,\gamma}$  (${\rm MeV cm^{-2} s^{-1}}$)} & \scriptsize{$(6.248\pm 0.899) \times10^{-5}$}\\
 \scriptsize{$\alpha_e$} & \scriptsize{$2.809 \pm 0.0520$}\\
 \scriptsize{$\epsilon^{syn}_{\gamma,c}$ (${\rm eV}$)} & \scriptsize{$0.100 \pm  0.001$}\\
 \scriptsize{$\epsilon^{syn}_{\gamma,m}$ (${\rm eV}$)} & \scriptsize{$(0.001 \pm 9.308)\times10^{-05}$}\\
\hline
 \end{tabular}
 &
 \begin{tabular}{cc}
 \hline
 \scriptsize{Parameter} & \scriptsize{Value}\\
 \hline 

\scriptsize{$A_{ssc,\gamma}$  (${\rm MeV cm^{-2} s^{-1}}$)} & \scriptsize{$(1.806\pm 0.754) \times10^{-5}$}\\
 \scriptsize{$\alpha_e$} & \scriptsize{$2.809 \pm 0.0520$}\\
 \scriptsize{$\epsilon^{ssc}_{\gamma,c}$ (${\rm MeV}$)} & \scriptsize{$7.254 \pm  2.574$}\\
 \scriptsize{$\epsilon^{ssc}_{\gamma,m}$ (${\rm keV}$)} & \scriptsize{$(11.374 \pm 3.696)$}\\
\hline 
\end{tabular}
\end{tabular}\\
\scriptsize{ \textbf{Table 1. The best fit of electron synchrotron radiation (left) and Compton scattering (right) parameters  obtained after fitting the SED.}} 
 

\end{document}